# Prediction of Ambient-Pressure High-Temperature Superconductivity in Doped Transition-Metal Hydrides

Haowei Xu [1], Olivia Schneble [2], Rafael Jaramillo [2], Marek Polański [3] and Ju Li [1,2,*]

[1] Department of Nuclear Science and Engineering, Massachusetts Institute of Technology, Cambridge, MA 02139, USA

[2] Department of Materials Science and Engineering, Massachusetts Institute of Technology, Cambridge, MA 02139, USA

[3] Military University of Technology, 2 Kaliskiego Street, 00-908 Warsaw, Poland

[*] Address correspondence to liju@mit.edu

## Abstract

The search for conventional superconductors with high transition temperatures ($T_c$) has largely focused on intrinsically metallic compounds. In this work, we explore the potential of intrinsically *non-metallic* compounds to exhibit high-$T_c$ superconductivity under ambient pressure through carrier doping. We identify $MgAlFeH_6$, a representative of carrier-doped transition-metal hydrides like $Mg_2FeH_6$, as a promising example with a predicted $T_c \approx 130$ K. We propose that the *average projected electron density of states*, defined as the geometric mean of the total and hydrogen-projected density of states at the Fermi level, serves as a simple and computationally inexpensive indicator of high-$T_c$ behavior. We also highlight the tradeoff between high-$T_c$ and dynamic stability, both of which depend on the electron density of states. Our findings thus expand the pool of potential superconducting materials and offer a practical route for accelerating the discovery of superconductors suitable for real-world applications.

Since the discovery of superconductivity in mercury in 1911, researchers have persistently sought materials with superconducting transition temperatures $T_c$ above room temperature ($\approx 300$ K), or at least above the temperature of liquid nitrogen ($\approx 77$ K). While the latter have been achieved in a class of cuprate compounds [1,2], their underlying physical mechanism remains elusive. In contrast, the mechanism of so-called conventional superconductivity is well established through the BCS theory [3,4], where phonons are responsible for electron pairing. For decades, the $T_c$ of conventional superconductors remains far below the temperature of liquid nitrogen, until the advent of hydride compounds, particularly $H_3S$, where a $T_c$ around 200 K has been observed in both *ab initio* calculations and experiments [5–7]. In addition, several other hydrides, including $YH_6$ [8], $CaH_6$ [9], and $LaH_{10}$ [10–12], have been proposed to have high $T_c$ as well.

Unfortunately, the hydride compounds mentioned above are dynamically stable only under extremely high pressures, typically exceeding 100 GPa, which severely limits their practical applicability. In this regard, recent research efforts have focused on discovering high-$T_c$ hydrides that can be stable under ambient pressure [13,14]. Particularly, using machine-learning accelerated high-throughput computation, Refs. [15,16] examined over a million compounds and predicted that a class of materials, $Mg_2XH_6$ with X = Rh, Ir, Pd, or Pt, could exhibit $T_c$ reaching 100 K at ambient pressure. From a practical perspective, however, these exciting results are undermined by the fact that Rh, Ir, Pd, and Pt are all noble metals with prices comparable to, or even exceeding, that of gold.

Clearly, it is a grand challenge to identify superconductors suitable for practical applications, namely those with low-cost, stability at ambient pressure, and high-$T_c$ above liquid nitrogen temperature or even room temperature. Actually, despite the vastness of the known materials database, which contains millions of entries [17,18], it is uncertain whether such materials really exist within it. It is worth noting that most previous studies have focused on intrinsically metallic materials, which limits the pool of candidates in materials genome searches. However, semiconductors and even insulators can have metallic behavior and exhibit superconductivity under certain conditions, such as carrier doping. Two prominent examples are the prototypical unconventional superconductor $YBa_2Cu_3O_{7-\delta}$ (YBCO), which is a Mott insulator in its undoped state ($\delta \approx 1$) [1,19], and boron-doped diamond, which becomes superconducting with a $T_c$ of approximately 4 K with sufficient boron dopants [20]. These findings motivate us to explore the potential of *non-metallic* materials, particularly hydrides, as superconductors, Especially when one consider the fact that multicomponent hydrides can exist as stoichiometric compounds but also it is very easy to partially substitute [21,22] elements changing the thermodynamical properties while keeping the crystal structure.

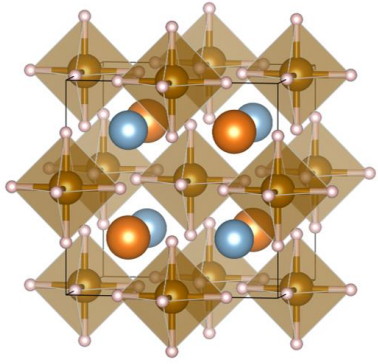

**Figure 1.** Atomic structure MgAlFeH$_6$ with an ordered lattice. Brown: Fe; orange: Mg; cyan: Al; pink: H. In pristine Mg$_2$FeH$_6$, both Mg and Al sites are occupied by Mg atoms.

In this work, we use ternary or quaternary transition-metal hydrides [23] as an example to demonstrate the potential of doped semiconductor/insulators as high-$T_c$ superconductors at ambient pressure. Instead of using brute-force high-throughput calculations, we adopt a rational materials design approach, beginning with non-metallic transition-metal hydrides as identified in Ref. [23]. Our initial focus is on compounds that exhibit van Hove singularities in the *average projected electronic density of states* (DOS) near the valence band maximum (VBM) or conduction band minimum (CBM). The average projected electron DOS is defined as the geometric average of the total DOS and the hydrogen-projected DOS. We argue that it can serve as an easy-to-compute indicator of potentially high superconducting critical temperatures ($T_c$). Carrier doping is then introduced through atomic substitution, insertion, or removal. The dynamic stability of the doped structures is assessed via their phonon band structures. For systems that are dynamically stable, we proceed to calculate their $T_c$. Our search is restricted to candidates with fewer than approximately 20 atoms per unit cell, which are tractable for *ab initio* $T_c$ calculations. Using this approach, we identify Mg$_2$FeH$_6$, which has been synthesized before [24–26] at ambient conditions by reactive mechanical alloying in a hydrogen atmosphere, and by high temperature solid-state reaction, as a promising platform for high-$T_c$ superconductivity at ambient pressure under carrier doping. In particular, we find that the electron-doped variant, MgAlFeH$_6$, exhibits a predicted $T_c$ of up to 130 K. Additionally, we highlight the critical interplay between high $T_c$ and dynamic stability, both of which can arise from a high electron DOS near the Fermi level. Our methodology, which emphasizes carrier doping in non-metallic systems, is not limited to transition-metal hydrides like $Mg_2FeH_6$. It can be broadly applied to expand the superconducting materials search space and potentially accelerate the discovery of compounds suitable for practical applications.

In the following, we first present the atomic structures, phonon spectra, and electronic properties of $Mg_2FeH_6$ and its electron- and hole-doped variants, $MgAlFeH_6$ and $MgNaFeH$. We find that $MgAlFeH_6$ is dynamically stable, whereas $MgNaFeH_6$ is unstable. We attribute this behavior to the exceptionally high electron DOS near the Fermi level in the latter. We then analyze the electron-phonon coupling and superconducting properties of $MgAlFeH_6$, highlighting a predicted $T_c$ of approximately 130 K. Finally, we explore the effect of charge doping on superconductivity in $Mg_2FeH_6$ in a more general context, demonstrating that $T_c$ up to 100 K can also be achieved. We also underscore a strong correlation between $T_c$ and the average projected electron DOS in doped $Mg_2FeH_6$.

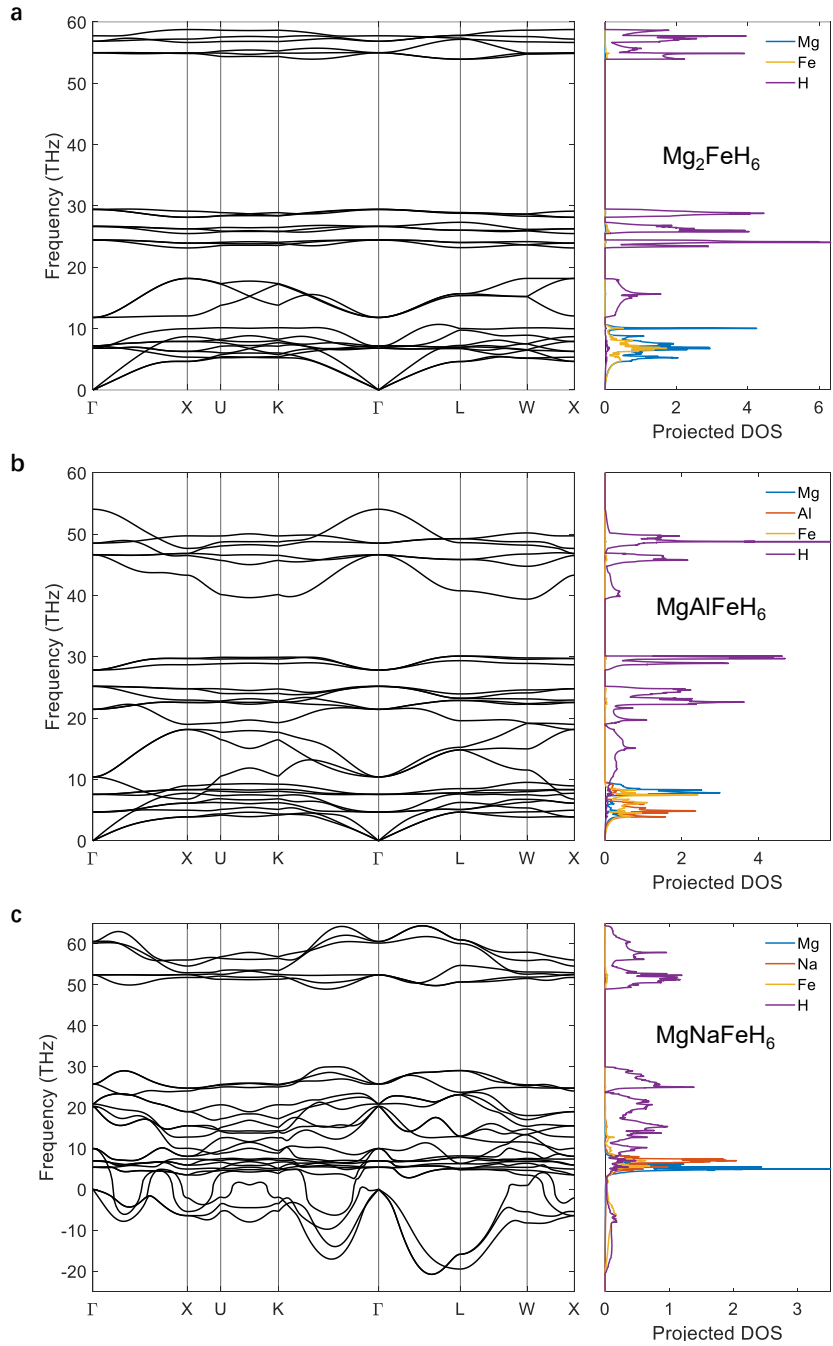

**Figure 2.** Phonon structure of (a) $Mg_2FeH_6$, (b) $MgAlFeH_6$, and (c) $MgNaFeH_6$. The phonon dispersion on high-symmetry lines in the Brillouin zone is plotted on the left-hand side, while the projected phonon DOS is plotted on the right-hand side. The DOS is in the unit of states per THz per formula unit. $Mg_2FeH_6$ and $MgAlFeH_6$ exhibit dynamic stability, while $MgNaFeH_6$ is dynamically unstable with imaginary frequency phonons spanning almost the entire Brillouin zone.

With 56 electrons per formula unit (f.u.), pristine $Mg_2FeH_6$ is a band insulator with a bandgap of around 2 eV, according to our density functional theory (DFT) calculations. It thus cannot be a superconductor in its intrinsic state. Nevertheless, the electron DOS of $Mg_2FeH_6$ features two van Hove peaks near the valence band maximum (VBM) and conduction band minimum (CBM). This property, together with the high-frequency phonons resulting from H atoms, suggests the potential of high $T_c$ if the electron Fermi level can be shifted into either the valence or the conduction bands. While this can be achieved by multiple strategies, such as removing or adding H atoms, we will focus on the effect of substituting Mg with Al (electron doping) or Na (hole doping) in this work. For now, we will explore the properties of $MgAlFeH_6$ ($MgNaFeH_6$) with an ordered lattice structure, that is, one of the two Mg atoms in the primitive cell is replaced by Al (Na). This is to keep the unit cell contains (9 atoms), so that *ab initio* calculations on $T_c$ can be tractable. Later, we will discuss how the electric properties of doped $Mg_2FeH_6$ depend on the carrier concentration under more general settings.

We begin by analyzing the atomic structure and dynamic stability of the Al/Na-substituted systems. Pristine $Mg_2FeH_6$ has a cubic structure with space group $Fm\bar{3}m$ (no. 225), which is reduced to $F\bar{4}3m$ (no. 216) after one of the two Mg atoms is replaced by Al or Na. Our DFT calculations indicate that $MgAlFeH_6$ tends to retain such a high symmetry structure, even if small perturbations are added to the initial structure before DFT relaxations. The conventional unit cell of $MgAlFeH_6$ is shown in Figure 1, where the lattice constant is around 6.24 Å. Mg, Al, and Fe atoms occupy the $4d$ $(0.75, 0.75, 0.75)$, $4c$ $(0.25, 0.25, 0.25)$, and $4b$ $(0.5, 0.5, 0.5)$ positions, respectively, while H atoms occupy the $24f$ $(0.24, 0, 0)$ Wyckoff positions. The phonon dispersions and projected phonon DOS of $Mg_2FeH_6$ and $MgAlFeH_6$ are plotted in Figure 2, showing similar characteristics. No imaginary frequencies are found over the whole Brillouin zone, indicating their dynamic stability. The low-frequency branches ($\lesssim 10$ THz) are mostly contributed by the metal atoms, while the high frequency branches ($10 \sim 60$ THz) are dominated by H atoms. The phonons of $MgAlFeH_6$ are slightly softened compared to $Mg_2FeH_6$. In contrast, $MgNaAlH_6$ turns out to be dynamically unstable, as its phonon dispersion exhibits imaginary frequencies over almost the entire Brillouin zone. We will return to this point later.

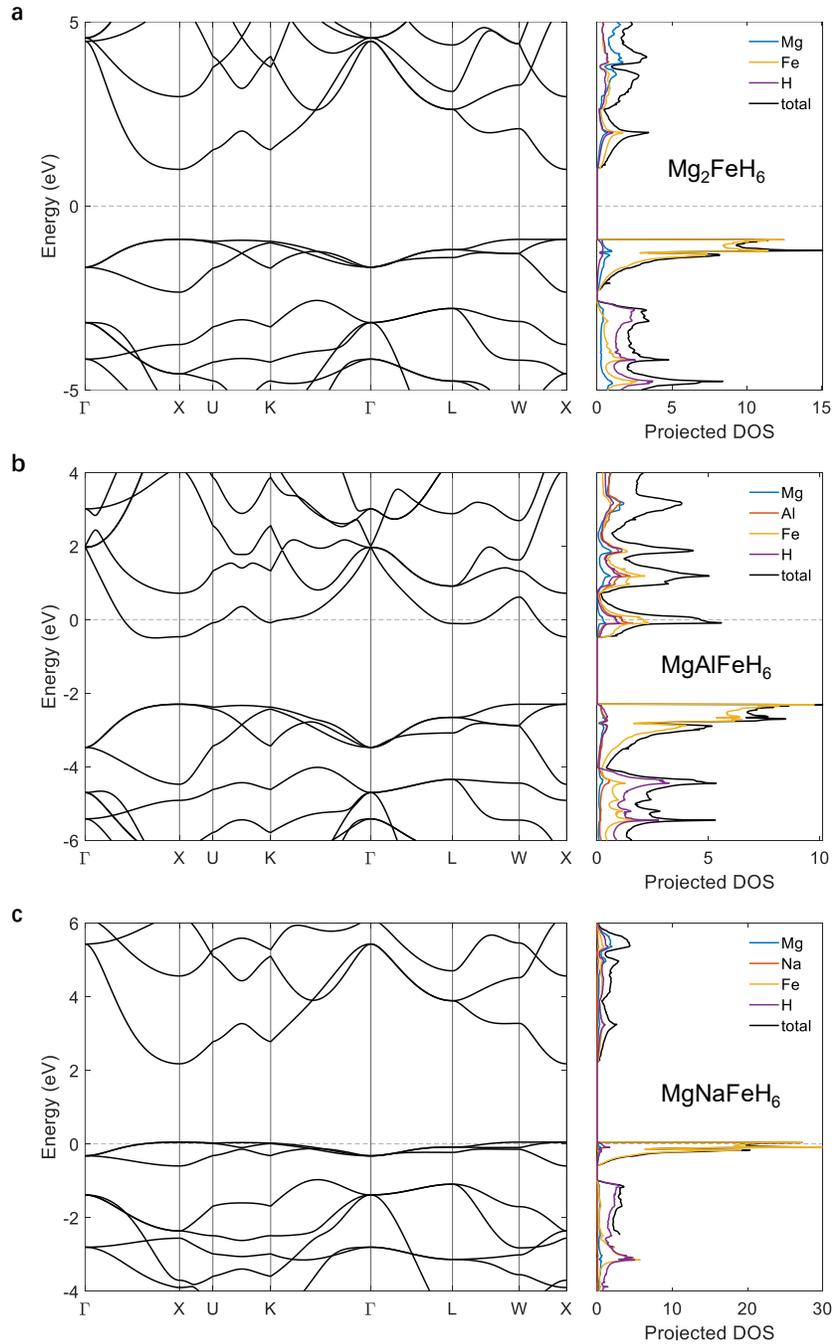

**Figure 3.** Electronic structure of (a) $Mg_2FeH_6$, (b) $MgAlFeH_6$, and (c) $MgNaFeH_6$. The electronic band structures on high-symmetry lines in the Brillouin zone are plotted on the left-hand side, while the projected electron DOS are plotted on the right-hand side. The DOS is expressed in units of states per eV per formula unit. The energy is offset to the electron Fermi level for each compound.

The electronic structures of MgAlFeH$_6$ and MgNaFeH$_6$ are plotted in Figure 3a and 3b, respectively. In contrast to Mg$_2$FeH$_6$, which has a bandgap of around 2 eV, the Fermi level of MgAlFeH$_6$ lies in the conduction band and is very close to the van Hove peak in the total electron DOS. Furthermore, the projected electron DOS indicates that H atoms have significant contributions to the total DOS, and the Fermi level is also close to the peak of the H-projected DOS. This suggests that when H atoms are displaced upon phonon excitation, the electronic structure may be significantly altered, potentially leading to strong electron-phonon coupling and high $T_c$.

On the other hand, the Fermi level of MgNaFeH$_6$ lies in the valence band and is close to an even more pronounced van Hove peak (Figure 3c), as the conduction bands are rather flat. Such an extremely high electron DOS at the Fermi level may be the reason why MgNaFeH$_6$ is dynamically unstable – the sharp and pronounced van Hove peak makes the electronic structure highly sensitive to atomic displacements [27]. This strong coupling between electronic and vibrational degrees of freedom may enhance electron-phonon interactions, which could favor a high $T_c$. However, the same mechanism also implies structural instability, as small perturbations may lead to drastic changes in the electronic properties, potentially destabilizing the system. This is indeed observed in the phonon band structure of MgNaFeH$_6$, as we discussed before. This highlights the critical tradeoff between high $T_c$ and dynamic stability in designing conventional (and potentially even unconventional) superconductors – both properties depend on the electron DOS near the Fermi level. While a high electron DOS enhances superconducting pairing, it may also destabilize the lattice, presenting a key challenge in materials optimization.

Next, we demonstrate the electron-phonon and superconducting properties in MgAlFeH$_6$. The Eliashberg spectral function $\alpha^2F(\omega)$ and the cumulative electron-phonon coupling (EPC) parameter $\lambda(\omega) = 2\int_0^\omega \frac{\alpha^2F(\omega)}{\omega}d\omega$ are plotted in Figure 4. The EPC constant $\lambda \equiv \lambda(\infty)$ is as large as 2.1, comparable to that of H$_3$S at 200 GPa [5–7]. The majority of the contributions to the EPC arise from the lower and middle regions of the phonon spectrum. To obtain the superconducting transition temperature $T_c$, we solve the anisotropic Eliashberg equations on the imaginary-axis and then continue to the real-axis by Padé approximation [28,29]. The histograms of the distribution of the anisotropic superconducting gap $\Delta_{n\mathbf{k}}$ is plotted in Figure 4b as a function of temperature $T$. The effective Coulomb potential $\mu_c$ in this calculation is an empirical parameter, which represents the repulsive electron-electron interaction. We used several different $\mu_c = 0.1, 0.125$, and $0.15$, which are typical values for conventional superconductors. By extrapolating the $\Delta_{n\mathbf{k}}(T)$ curves to $\Delta_{n\mathbf{k}}(T_c) = 0$, we obtain $T_c \approx 130$ K.

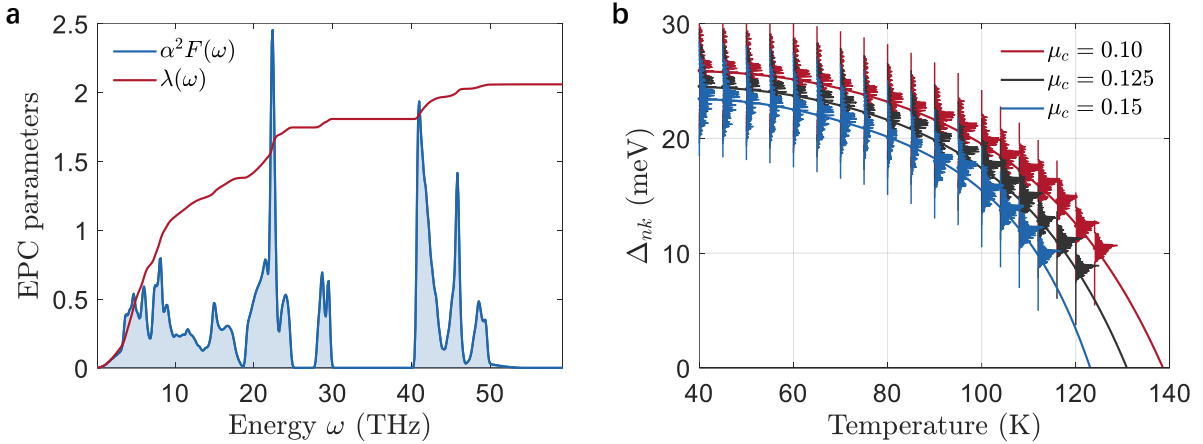

**Figure 4.** Electron-phonon coupling and superconductivity of MgAlFeH$_6$. (a) Electron-phonon coupling (EPC) parameters $\alpha^2 F(\omega)$ and $\lambda(\omega)$. The EPC constant is $\lambda(\omega \to \infty) \approx 2.1$. The effective Colomb potential is set to $\mu_c = 0.1$ for this plot. (b) Histograms of the distribution of the anisotropic superconducting gap $\Delta_{n\mathbf{k}}$ as a function of energy, which are obtained by solving the anisotropic Eliashberg equation at different temperatures $T$. Three different $\mu_c = 0.1, 0.125$, and $0.15$ are used. The solid lines fit the average of $\Delta_{n\mathbf{k}}$ at each temperature, which are then extrapolated to yield the superconducting transition temperature $T_c$.

The above discussions assumed an ordered lattice of MgAlFeH$_6$. In practice, depending on the synthesis approach, Mg and Al atoms may randomly occupy the original sites of Mg atoms in pristine Mg$_2$FeH$_6$. To account for this effect, we constructed a supercell Mg$_{32}$Al$_{32}$Fe$_{32}$H$_{192}$ with 288 atoms in total. Here, 32 Mg and 32 Al atoms are randomly placed, representing a random alloy. Then, we calculate the electronic and phonon properties of this supercell, which are shown in Figure 5. This structure is dynamically stable with no imaginary frequency phonons. Compared with ordered MgAlFeH$_6$, the electron and phonon DOS are smeared out due to inhomogeneity. Nevertheless, some characteristic features that prelude high $T_c$ remain. Particularly, the electron Fermi level is located near the peak of the electron DOS, where H atoms make significant contributions. As for phonon properties, metal atoms contribute mostly to low frequency phonon modes $\lesssim 10$ THz, while high frequency phonon modes ($10 \sim 60$ THz) are dominated by the vibrations of H atoms. These features resemble those of ordered MgAlFeH$_6$. Hence, a similarly high $T_c$ could be expected, though a precise calculation of $T_c$ of this supercell is not possible due to the formidable computational cost.

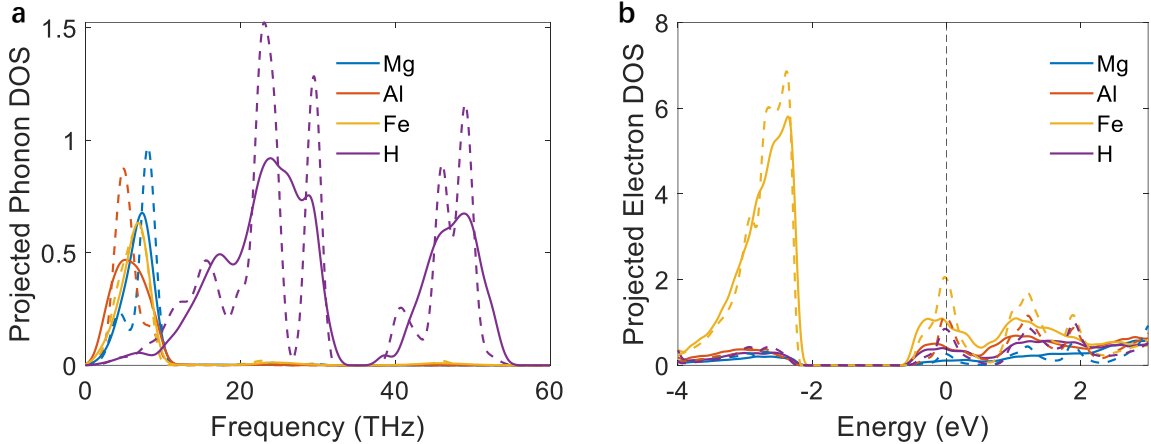

**Figure 5.** The solid curves are projected (a) phonon and (b) electron DOS of a Mg$_{32}$Al$_{32}$Fe$_{32}$H$_{192}$ supercell. Mg and Al are randomly placed at the Mg sites in pristine Mg$_2$FeH$_6$, which represents a random alloy. The dashed curves are results for ordered MgAlFeH$_6$. In (b), the energy is offset to the electron Fermi level. The units for the y-axis of (a) / (b) are states per THz / eV per MgAlFeH$_6$ formula unit.

As previously mentioned, substituting Mg with Al is just one of several possible strategies for inducing superconductivity in Mg$_2$FeH$_6$. In principle, any method that introduces free carriers could be effective. To explore these strategies in a more general way, we directly vary the total number of electrons in Mg$_2$FeH$_6$ within our DFT calculations. Reducing (increasing) the electron number is equivalent to hole (electron) doping. Note that varying the total number of electrons leads to a divergent Coulomb energy, which can be compensated by adding a uniform background charge of opposite sign in DFT calculations. We adopt this approach to calculate the superconducting critical temperature $T_c$ of charge-doped Mg$_2$FeH$_6$ as a function of doping concentration $n$, where positive and negative values of $n$ (in units of electrons per formula unit, $e$/f.u., equivalent to $1.6 \times 10^{22}$ cm$^{-3}$) correspond to hole and electron doping, respectively. When $n \lesssim \pm 0.1\ e/$f.u., the predicted $T_c$ is negligibly small. This behavior is consistent with experimental observations in boron-doped diamond, where the system remains semiconducting below a similar doping threshold [20]. For $n \gtrsim \pm 0.1\ e/$f.u., $T_c$ becomes appreciable and generally increases with doping concentration. At $n \approx \pm 1\ e/$f.u. – equivalent to the carrier concentrations in MgAlFeH$_6$ and MgNaFeH$_6$ – $T_c$ reaches approximately 100 K, in agreement with the directly calculated $T_c$ for MgAlFeH$_6$. These results support the idea that carrier doping in semiconductors is a viable route to achieving high-$T_c$ superconductivity under ambient conditions. Note that while $T_c$ tends to continue increasing with $n \gtrsim +1\ e/\text{f.u.}$, the structure becomes dynamically stable in this regime. This again highlights the critical tradeoff between high $T_c$ and dynamic stability, both of which depend on the electron DOS. Here it is worth mentioning that the MgAlFeH$_6$ compound studied before is an alloy of the parent compound, but that the core concept can be seen at lower electron concentrations, down towards the regime of doping.

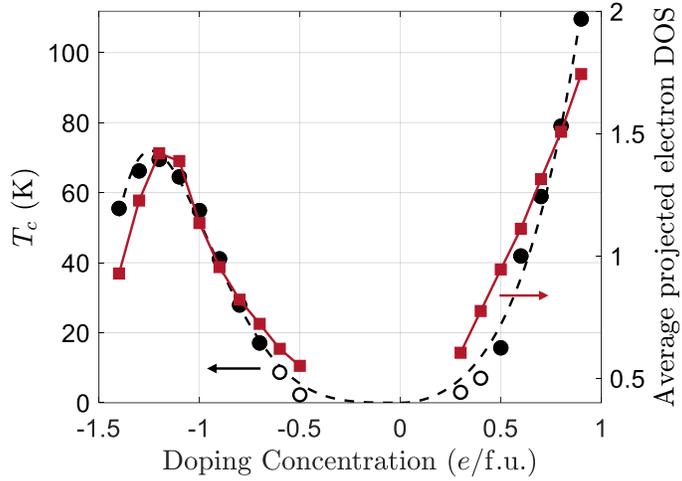

**Figure 6.** Left $y$-axis: superconducting transition temperature $T_c$ as a the function of carrier doping concentration in Mg$_2$FeH$_6$. Filled black circles represent values obtained by solving the anisotropic Eliashberg equations, while open black circles are deduced from the Allen-Dynes equation. The dashed black curve serves as a guide to the eye. Right $y$-axis: the red curve is the average projected electron DOS in the unit of states per eV per formula unit.

Here, a critical observation is that the predicted $T_c$ shows a strong correlation with the average projected electron DOS at the Fermi level, defined as $\bar{D} = \sqrt{D_{\text{total}} D_{\text{H}}}$, where $D_{\text{total}}$ and $D_{\text{H}}$ are the total and H-projected electron DOS at the Fermi level, respectively. $T_c$ and $\bar{D}$ have almost the same trend as a function of doping concentration. This phenomenological correlation can be understood by noting that $D_{\text{total}}$ reflects the number of electrons available for superconducting pairing, while $D_{\text{H}}$ is closely related to the strength of electron-phonon coupling, which is primarily mediated by H atoms in hydrides. Compared with phonon and electron-phonon coupling properties, $D_{\text{total}}$ and $D_\text{H}$ are purely electronic quantities that can be efficiently computed from *ab initio* calculations with relatively low computational cost. As such, $\bar{D}$ may serve as a convenient and practical descriptor for guiding the search for high-$T_c$ superconductivity in both metallic and doped non-metallic hydrides – one can simply look for systems where $\bar{D}$ is maximized at the Fermi level, typically near a van Hove singularity. For compounds lacking hydrogen, $\bar{D}$ can be generalized as $\bar{D} = \sqrt{D_\text{total} D_\text{X}}$, where X refers to other light atoms, such as boron.

From an experimental perspective, electron- and hole-doping may be achieved electrically, electrochemically, or chemically. Electrical doping, or charge doping, can be achieved by applying an electric field or by creating a heterojunction, in which a local space charge is induced. Such controllable and localized (nearly 2D) superconductivity may be of interest for numerous applications in information processing or sensing. The 3D substrate may also provide mechanical constraint to prevent phase transformation that may otherwise would happen. To carry a lot of current, however, requires 3D superconductors, and this may be achieved by electrochemical hydrogenation [30], chemical hydrogenation, chemical alloying or chemical substitution, where electroneutrality is satisfied in the bulk. Generally speaking, n- or electron doping corresponds to decreasing the electrochemical voltage $U$ or going to more "reducing" local chemical conditions (e.g. lowering the oxygen partial pressure $P_{O2}$ or increasing the hydrogen partial pressure $P_\text{H2}$), whereas p- or hole doping corresponds to increasing the electrochemical voltage $U$ or going to more oxidative local chemical conditions (e.g. $P_{O2} \uparrow P_\text{H2} \downarrow$), as the typical Brouwer diagram of semiconducting oxide shows. We note that even though MgAlFeH$_6$, MgNaFeH$_6$, etc. appear to have unusual valences, such exotic-valence compounds do exist in nature, for example high-voltage cathode oxide particles inside lithium-ion batteries [31], or zirconium suboxide (ZrO) in metallic zirconium oxidation [32]. Achieving these exotic-valence compounds by going to very reducing or oxidative local conditions either thermodynamically or kinetically [33,34] may be still experimentally more feasible than going to hundreds of GPa mechanical pressure.

In summary, we demonstrate that Mg$_2$FeH$_6$, an intrinsic semiconductor, can be transformed into an ambient-pressure superconductor with $T_c$ above 100 K through carrier doping. Unlike most previous studies that have focused on intrinsically metallic compounds, our work highlights the potential of non-metallic systems as promising platforms for high-temperature superconductivity upon carrier doping, thereby broadening the pool of candidate materials. Furthermore, we identify the average projected electron density of states as an easy-to-compute genome that can accelerate the discovery of high-$T_c$ superconductors. Future work applying these methodologies may lead to the discovery of additional superconductors suitable for practical applications.

## Methods

The first-principles calculations in this work are based on density functional theory (DFT) [35,36] using the Quantum Espresso package [37,38]. The PBEsol [39,40] exchange-correlation functional is adopted. Core and valence electrons are treated with the projected augmented wave (PAW) method [41] and plane wave basis, respectively. Electronic Hamiltonians in the Wannier basis are then built from plane-wave DFT results, using the Wannier90 package [42]. The electron-phonon coupling and superconductivity properties are calculated with the help of the EPW package [43,44]. To compute the superconducting transition temperature, $5 \times 5 \times 5$ coarse (before Wannier interpolation) and $25 \times 25 \times 25$ fine (after Wannier interpolation) meshes are used to sample the first Brillouin zone of both electrons and phonons.

## Acknowledgments.

We acknowledge support by the Gordon and Betty Moore Foundation's EPiQS Initiative, Grant GBMF11945.